\documentstyle[12pt,epsf,epsfig]{article}
\def\lsim{\lower.5ex\hbox{$\; \buildrel < \over \sim \;$}}
\def\gsim{\lower.5ex\hbox{$\; \buildrel > \over \sim \;$}}

\def\b1#1{\hbox{$^{1#1}{\rm B}$}}

\def\msun{${\,M_\odot}$}
\def\etal{{\it et al.}~}
\def\beginapjbib{\begingroup \section*{\large \bf References}
   \parskip=.5ex plus 1.0pt
   \def\bibitem{\par \noindent \hangindent\parindent
      \hangafter=1}}
\def\endapjbib{\par \endgroup}

\def\beq{\begin{equation}}
\def\eeq{\end{equation}}

\begin{document}
\begin{titlepage}
\pagestyle{empty}

\rightline{UMN-TH-1615/97}
\rightline{astro-ph/9712261}
\rightline{November 1997}
\vskip .2in
\begin{center}
{\large{\bf Star Formation Histories versus Redshift :
 Consequences for Overall Metallicity and Deuterium Destruction\footnote{New
Astronomy, in press}}}
\end{center}

\begin{center}
Michel Cass\'{e}$^1$,
Keith A. Olive$^2$, \\
Elisabeth Vangioni-Flam$^3$
 and Jean Audouze$^3$

$^1${\it Service d'Astrophysique, DSM, DAPNIA, CEA,  France}

$^2${\it School of Physics and Astronomy, University of Minnesota \\
 Minneapolis, MN 55455, USA}

$^3${\it Institut d'Astrophysique de Paris, 98bis
Boulevard Arago, \\ 75014 Paris, France}

\vskip .1in

\end{center}
\vskip .3in
\centerline{ {\bf Abstract} }
\baselineskip=18pt

The flood of new data on deep surveys, and above all the CFRS (Canada-France
 -Redshift-Survey), has had
a great impact on studies of galactic evolution. 
On the basis of cosmological
models consistent with the improved values of the
 Hubble parameter,
 different star formation histories are tested against 
the observed UV, B and IR
 broad band comoving luminosity densities. Using these
spectrophotometric results, we analyze the
 global metal enrichment with the help of chemical 
evolutionary models and we discuss the pertinence of different metallicity
tracers ( quasar absorption
 systems and clusters of galaxies) as representative of the bulk
 chemical evolution of the Universe. 
Moreover, as deuterium is  very fragile, this isotope is destroyed 
in all stars and its evolution
 is particularly sensitive to the history of star formation. 
Relying on models constrained to fit the solar vicinity, it is shown
that models with high D destruction corresponding to a large decrease of
 the star formation rate (SFR) from $z=1.5$ to 0 are in good agreement with
 spectrophotometric data. In contrast, low D destruction models
 which require only a moderate variation of the SFR in the same redshift
range
 seem to encounter difficulties in matching the evolution of the luminosity
 densities (UV, B and IR) versus redshift. The sensitivity 
 of the results with the cosmological models of the universe
 is discussed.


\baselineskip=18pt

\noindent
\end{titlepage}
\baselineskip=18pt                   

\section{Introduction}

 UV, optical, and IR observations of high redshift 
objects have recently achieved
 a spectacular breakthrough (Lilly \etal 1996, Steidel \etal 1996,
 Sawicki , Lin, \& Yee 1997, Madau \etal 1996, Treyer et al 1997),
revealing a relatively intense period of star formation at 1$<z<$2 (Madau
\etal 1996, 1997) with a decrease by a factor of about ten since then. 
The Canada-France Redshift Survey (CFRS) has produced data on the
comoving luminosity densities ${\cal L}(\lambda, z)$ in three wavelength
bands (0.28, 0.44, and 1 $\mu$m) over the redshift range $0 < z < 1$
(Lilly \etal 1996). These data are complemented by the Hubble Deep Field
(HDF) data in the redshift range $0.5 < z < 2$ at 0.28 $\mu$m
(Connolly \etal 1997) and in the range $0.2 < z < 4$ at 0.3 $\mu$m and
0.45 $\mu$m (Sawicki, Lin, \& Yee 1997) . Because it is expected that the
UV luminosity density of star forming galaxies is related to their
star-formation rate, these data have allowed one to map out the cosmic
star formation history as well as the metal enrichment history as a
function of redshift out to
$z \approx 2$.
Note however, that above $z = 2$ the luminosity data are still
uncertain. Specifically, the luminosity function of Lilly \etal (1996) are the
 most reliable being derived from spectroscopic redshifts plus photometric data.
 The luminosity functions of Connolly \etal (1997) and Sawicki \etal (1997)
  are based on photometric redshifts, and are subject to uncertainties. This
  is in addition to the uncertainties due to dust at high redshift. We stress that
 the lower redshift data provide significant constraints on the models considered,
thus our essential conclusions are not strongly dependent on the high
 redshift data.

Indeed, because of these observations, it has become possible to 
extend the earlier notions of so-called cosmic chemical
evolution (see e.g. Pei \& Fall 1995) based on damped Ly$\alpha$
systems (DLAs). Such systems show a large variation of comoving H I with
respect to redshift (Lanzetta \etal 1995) and a metallicity of about
 Z $\approx 0.1 $Z$_\odot$ at $z \sim 2$, however with a large dispersion
 (Pettini \etal
1997c). Although, as we will argue below, these systems may not be
representative of the globally averaged star formation history, 
since they map out the outer part of spiral
galaxies (Phillips \& Edmunds 1996) and/or protogalactic clumps not
 totally constituted in galaxies.  In
addition, it was shown  (Fall \& Pei 1993; Pei
\& Fall 1995) that obscuration plays a crucial role in understanding the
evolution of such systems particularly at high redshift.  Because of the
problems associated with obscuration, observations at somewhat lower
redshift $z \lsim 2$ are interesting from the point of view of cosmic
chemical evolution and allow one to model the bulk of the 
history for galaxy evolution.

It is worth noting however, that  the conversion of the luminosity
densities into a star formation rate (SFR) is model dependent through
the adopted initial mass function (IMF).  In addition, the corrections
needed to account for dust obscuration are rather uncertain and most
probably lead to an upward revision of the star formation 
rate at high redshift by a
factor of a few deduced from UV observations (Pettini \etal 1997b). This
 seriously affects the high redshift observation at 0.15 $\mu$m by
the HDF (Madau
\etal 1996) and we treat this data as a lower limit on the
luminosity density at that wavelength. With the same caveat, we also
 consider the HDF data of Sawicki, Lin, \& Yee 1997, which give higher
 values. As we said, obscuration
 is expected to be small at low $z$, since we know that only a third of
 the luminosity of galaxies at $z=0$ is radiated in the IR (Soifer \&
 Neugebauer 1991). Thus, the CFRS data (Lilly \etal 1996) are of special
 interest.  

Independently, the evolution of the overall
metallicity Z, can be derived from observations of 
heavy abundances seen in i) 
absorbing systems along the line of sight to distant quasars
(DLAs, usually
associated with very young galaxies, Pettini \etal 1997a,
 Lu \etal 1996, 1997); ii) Lyman$\alpha$ forest 
(associated with intergalactic gas clouds, e.g. Savaglio 1997)
iii) X-ray emitting galactic clusters associated with the intracluster gas
 (Mushotzky \& Loewenstein 1997,
Renzini 1997).  Though these observations may serve as an additional
constraint on cosmic star formation histories, they
 have to be carefully interpreted to determine
what medium is the most representative of the history of star formation
as revealed by the spectrophotometric data.
Phillips \& Edmunds (1996) and Edmunds \& Phillips (1997) have considered in detail
the cosmic chemical  evolution following the evolution of the mean metal abundance
of the
 Universe. They have taken into account all the different classes of galaxies
 and their internal components. This is certainly the best way to treat the
 problem though there are many obstacles. In this paper, our main goal is
 to confront the D history with the luminosity density  changes as a function
 of the redshift. The observations of the present or near present D/H ratio
 are only available in the solar vicinity (ISM and solar system).
 Thus, the galactic evolutionary models selected are anchored to the local
 galactic environment. Even if high redshift estimates exist, they are 
unfortunatly quite
 uncertain.  It is clear that when we will have a broader data set at our disposal,
some of these questions can be reassessed. The  FUSE satellite developed by NASA in
collaboration with France and Canada will be launched at the end of 1998. One
  of its principle goals is to measure D in various places: the galactic center and
 anti-center, the galactic halo and external galactic sites (Vidal-Madjar et al
 1997). Though we are making use of models designed for only a part of the Galaxy,
whereas the luminosity
 data are  representative of an average over various parts and types of
 galaxies, it is edifying to confront these models to global constraints at
 moderate redshift to test, among other things, their local or general
 character.
It turns out that certain models, among the ones studied, lead to a good
 reproduction of the luminosity density in different wave bands; this
 convergence is obtained with models similar to empirical ones by Madau (1997).
 This points to a large variation of the SFR from $z=2$ to $z=0$ in
 agreement with  Edmunds \& Phillips (1997).

In the context of cosmic chemical evolution, the work  by Madau (1997),
 Madau \etal
(1997) is quite illustrative. They have considered three different IMFs: a
Salpeter IMF with slope
$x = 1.35$,  $x = 1.7$ IMF and a Scalo (1986) IMF, the latter
yielding a poor fit to the data at the larger wavelengths due to an
excess of approximately solar mass stars. In agreement with Lilly \etal
(1996), the data are fit better with an IMF which favors the
flatter Salpeter IMF. Madau
\etal (1997) also considered two different star formation rates, both
of which increase from
$z=0$ to $z =1.5$. Of the two,  one decreases at $z \gsim 1.5$,
 while the second remains almost constant at higher redshift. 
The latter requires dust obscuration which increases with redshift.
  From the observations of the
DLAs alone, they favor the first mode. Such a 
 correction may not be necessary on the basis of the HDF data of Sawicki,
Lin, \& Yee (1997) which indicates a relatively flat luminosity density 
 between $z$ = 1 to about 4.

In what follows, we reassess the problem of tracing the cosmic
chemical evolution with the available high redshift data, using the
 population synthesis model of Bruzual and Charlot (1997). As such, we
are presented with a good 
 occasion to enlarge the debate on galactic evolution and promote 
 local arguments to cosmological ones. We will
consider several very different star formation rate histories
including the relevant  metallicity information. We find that the high
redshift data is indeed very sensitive to the form of the adopted SFR.

The star formation rate in a chemical evolution model has a strong impact
on the degree to which deuterium is destroyed.  Given a definite
primordial value for D/H, the present day D/H abundance is a prediction
of the model.  Indeed different models of galactic chemical evolution
make very different predictions concerning the total amount of deuterium
astration.
Given the current uncertainties in
the observations which correspond to the primordial abundance of D/H,
 several different SFRs have been employed for suitably
describing the evolution of D whether or not the primordial abundance
of D/H was high or low. We will consider the consequences of
models (Scully \etal
1997, Timmes, Woosley, \& Weaver 1995, Tosi 1996 and references therein)
corresponding to both high and low D destruction (see table 1),
 on the comoving luminosity densities in three bands
${\cal L}(\lambda, z)$ upon extrapolating them to a  scale
 representative of the Universe as a whole. 

To this end, we will briefly review the photometric and chemical
evolution models considered in section 2. In section 3, we 
 display our results for the models
considered and compare these to the data. In section 4, we discuss the 
consequences of these models on the destruction of Deuterium and
 the metal enrichment history at high
redshift. 
Finally, in the last section 
we present our conclusions and perspectives.
 
\section{Star formation histories in chemical evolution 
models and population synthesis}

Simple models of galactic chemical evolution have enjoyed considerable
success in fitting abundance data in the Galaxy and solar neighborhood
(Tinsley 1980). To first order, such models require some form for the
IMF and SFR as broad averaged quantities. When combined with element
abundance yields from supernovae, a simple integration involving the
gas mass and yields over the IMF and SFR, gives abundance ratios
and total abundances of heavy elements in  general agreement with
observations. However, unless one has considerable faith in these simple
models, it is difficult to make predictions.  For example, many
simple models of chemical evolution, as well as some more complicated
ones which include the infall of matter onto the galactic disk, often
``predict" that the total amount of deuterium destruction that has
occurred in the solar neighborhood is limited to a factor of 2 -- 3 (Tosi 1996).
However, it has been shown (Scully \etal 1997) that by including the
effects of  outflow or galactic winds driven by supernovae heating in the
early galaxy, the amount of deuterium destruction can be significantly
larger when coupled to a steeply decreasing SFR.
That such models exist is clearly of importance if the value of 
D/H in quasar absorption systems are as high as some observations
indicate (Songaila \etal 1994, 1997, Carswell \etal 1994, Wampler 
1996, Webb \etal 1997).

The chemical evolution models of Scully \etal (1997) were designed to
compare different degrees of deuterium destruction. As such, the models
were constructed to first match solar and present day D/H values
of (D/H)$_\odot \simeq 2.6 \times 10^{-5}$ (Geiss 1993, Scully \etal 1996)
and D/H$_{\rm ISM} \simeq 1.6 \times 10^{-5}$ (Linsky \etal 1993, 1995)
to the high D/H measurements in quasar absorption systems (Songaila \etal
1994, Carswell \etal 1994, Webb \etal 1997) of $\sim 2 \times 10^{-4}$;
the low D/H measurements of $2.5 \times 10^{-5}$ 
(Tytler, Fan, \& Burles 1996, Burles \& Tytler 1996); and an intermediate
value of D/H $ = 7.5 \times 10^{-5}$. Parameters of the models were then
adjusted so as to fit the abundance ratios [Fe/H] vs. $t$ and [O/H] vs.
$t$, the present day gas mass fraction, and G-dwarf distribution vs.
metallicity in the solar neighborhood.

Because of the increased amount of stellar processing needed to destroy
deuterium in the high primordial D/H cases, galactic winds were introduced
to avoid the overproduction of heavy elements. The winds generally
included two components (Vader 1986), one which is due to the heating of
the ISM from the dissipation of energy of the hot supernova remnants
(Larson 1974) and a second component which is metal enhanced and is
directly proportional to the supernovae rate with an efficiency $\nu$.
In all of the models, a single slope IMF with $x = 1.7$ was used.
For the high D/H values, two models were constructed. In model 1a, 
a bimodal model of star formation (Larson 1986) was considered with
a massive mode (of stars with masses between 2 -- 100 \msun) with an
exponentially decreasing SFR, $\psi = 0.29 e^{-t/2}$ (times in Gyr)
and a ``normal" mode (stars with masses between 0.4 -- 100 \msun)
with a SFR given by $\psi = 0.29 M_G$, where $M_G$ is the mass in gas.
In this case $\nu$ was set to 0.81. A second model was constructed for
the high D/H case which better matched the local G-Dwarf distribution, 
because it models a prompt initial enrichment. This case, called model
2, is a sequential model in which the massive mode has a SFR given by
$\psi = 0.19 e^{-t/1}$ for
$t \le 1$ Gyr, afterwhich it is replaced by the normal mode with
 $\psi = 0.73 e^{-t/2.5}$. The IMF is the same as in 1a and here $\nu =
0.68$.

Models 1b and 1c were designed to match the D/H evolution with
primordial D/H = 7.5 $\times 10^{-5}$ and 2.5 $\times 10^{-5}$
respectively. In these models a single SFR was used. For model 1b, $\psi
= 0.28 M_G$ with the normal mode IMF of models 1a and 2 and $\nu = 0.55$,
and in Model 1c, $\psi = 0.07$, i.e., a constant SFR with a normal mode
IMF which extended down to 0.2 \msun.  In this case, since metal
production rather than overproduction is a problem, one sets $\nu =
0$.

It  is also of interest here to compare our results based on the above
models along with   galactic chemical evolution models which use infall to
 progressively form the galactic disk.
We will therefore consider two models of this type: the model of 
Timmes, Woosley, \& Weaver (1995) hereafter TWW; and a model
representative of the compilation in Tosi (1996) which includes models of
Carigi (1996), Galli \etal (1994); Matteucci \& Francois (1989); and
Prantzos (1996).  The latter set of models, all have similar star
formation histories as well as D/H evolution.  They differ more in their
galactic radial dependence which does not concern us here. All of these
models including that of TWW (1995) all employ infall and they are
dependent on the gas density, $\sigma$. While the model of TWW has a
relatively strong dependence ($\sigma^2$), the others have a more
moderate dependence (closer to a linear dependence as in models 1a,b
above). 

In Figure 1, we show all of the SFRs (as a function of redshift) of the
models considered. The various curves are labeled 1a,b,c and 2
corresponding to the models of Scully \etal (1997), TWW, and T
corresponding to a SFR representative of the models taken from the
compilation of Tosi (1996).  For comparison, we also show the two star
formation rates from Madau \etal (1997).  They have all been normalized  
at $z = 0$ so that log(SFR) = 0.

In what follows in the next section, we will assume that the adopted  star
formation history is not too different from the history
in an average galaxy which contributes to the luminosity density at
high redshift. In fact, we know from the work of Madau \etal (1997)
that this is not too bad an assumption since the models considered
there with respect to the high redshift data resemble standard
chemical evolution models.  Therefore we take each of the models
above, assume that the various IMFs and SFRs are in fact universal and
extract the luminosity density as a function of time (or redshift for
a given cosmological model) as described below.  This luminosity
function is then compared to the available data.

We compute the spectrophotometric properties of model galaxies using new
population synthesis models by Bruzual \& Charlot (1997). These span the
range of metallicities $5\times10^{-3}\leq Z/Z_\odot \leq 5$ and include
all phases of stellar evolution from the zero-age main sequence to supernova
explosions for progenitors more massive than $8\,M_{\odot}$, or the end of
the white dwarf cooling sequence for less massive progenitors. The models
are based on recent stellar evolutionary tracks computed by Alongi \etal
(1993), Bressan \etal (1993), Fagotto \etal (1994a, b, c), and Girardi
\etal (1996), supplemented with prescriptions for upper-AGB and post-AGB
evolution. The radiative opacities are taken from Iglesias \etal
(1992).  In the version used here, we adopt the library of synthetic
stellar spectra  compiled by Lejeune \etal (1997) for all metallicities.
This library is based on spectra by Kurucz (1995, private communication;
see also Kurucz 1992) for the hotter (O-K) stars, Bessell \etal (1989,
1991) and Fluks \etal (1994) for M giants, and Allard \& Hauschildt
(1995) for M dwarfs. The Lejeune \etal spectral library also includes
semi-empirical corrections for blanketing, a well-known limitation of
synthetic spectra (see, for example, Charlot, Worthey \& Bressan 1996,
and references therein). The resulting model spectra computed for stellar
populations of various ages and metallicities have been checked against
observed spectra of star clusters and galaxies (Bruzual \& Charlot 1997;
Bruzual \etal 1997).

A complete discussion of the differences between the spectrophotometric
predictions of these models with those in previous studies will be presented
in Bruzual \& Charlot (1997).  The typical discrepancies between the properties
of stellar populations of the same input age and metallicity that are obtained
by using the spectral synthesis models constructed by different groups
of scientists, have already been illustrated by Charlot, Worthey \& Bressan
(1996). These can reach up to 0.05~mag in rest-frame $B-V$, 0.25~mag in
rest-frame $V-K$ and a 25\% dispersion in the $V$-band mass-to-light ratio.
With these uncertainties in mind, we will concentrate more on understanding
the trends seen in the observations than on obtaining exact fits to the data.

Finally, since our SFRs are all expressed in terms of time, we need a
cosmological model to convert time to redshift.  The galactic
evolution models described above were originally designed to model
our Galaxy with a total age 14 Gyr, a shorter time would require a
more rapidly changing SFR, particularly for models 1a and 2. The
conversion of course is well known
\beq
H_o t = \int_0^{(1+z)^{-1}} {dx \over \sqrt{1 - \Omega_o - \Omega_\Lambda
 + \Omega_\Lambda x^2  + \Omega_o/x} }
\label{t}
\eeq
where $\Omega_\Lambda = \Lambda/3H^2$ and $q_o = \Omega_o/2 -
\Omega_\Lambda$. For the simple case of an Einstein-de Sitter Universe
($q_o = 1/2$ and $\Lambda = 0$), the right hand side of Eq. (\ref{t}) is
just ${2
\over 3} (1 + z)^{-3/2}$. For comparison we will also consider a $q_o =
0.1$ Universe as well.

\section {Results confronted to photometric observations}

As indicated in section 1, we compare the results of our model
calculations to the high redshift data of Lilly \etal  (1996),
Connolly \etal (1997), and Sawicki \etal (1997). Lilly
\etal used the CFRS galaxy sample for which both redshift and
B,V,K and I band photometry was available. As such, they constructed
the co-moving luminosity densities ${\cal L}(\lambda,z)$ in three
redshift bins  ($z$ = 0.2 -- 0.5,  0.5 -- 0.75, and 0.75 -- 1.0) and
at three wavelengths (0.28, 0.44, and 1.0
$\mu$m).  For the local value ($z = 0$) they adopted the value 
from Loveday  \etal (1992) and we will do the same.  (We note that
there  has been considerable discussion in the literature recently, 
regarding the local value).
We also compare our results to the data of Connolly \etal (1997) based
on HDF data at 0.28 $\mu$m in the redshift ranges 0.5 -- 1.0,
1.0 -- 1.5, 1.5 -- 2.0.
The HDF data of Madau \etal (1996) at 0.15 at $z > 2$, is difficult
to use for our purpose here because of the effects of extinction
which may range from a factor of $\sim$ 3
(Pettini \etal 1997b) to as much as a factor of $\sim$ 15 (Meurer 1997).
For completeness, we also compare our results with the HDF data of 
Sawicki \etal (1997) which are at the slightly different wavelengths
 of 0.3 and 0.45 $\mu$m (we will ignore this difference with the other
data) and in the redshift bins, 0.2 -- 0.5, 0.5 -- 1.0, 1 -- 2, 
2 -- 3, and 3 -- 4.

As we discussed in section 2, one of our main goals in this work is to
test the various star formation histories employed in several galactic
chemical evolution models against the high redshift luminosity density.
The SFRs considered are indeed very different
and range from a constant SFR which is flat as a function of $z$, to one
that is a steeply decreasing exponential with time.  These are shown in
Figure 1. When run through the population synthesis code (Bruzual and
Charlot 1997), it will be clear that the high redshift luminosity data
is very sensitive to the input SFR.

In Figures 2 - 7, we show the data taken from Lilly \etal (1996)
shown as filled squares for $\lambda$ = 0.28, 0.44, and 1.0 $\mu$m, and
from Connolly \etal (1997) shown as open circles for the 0.28 $\mu$m
wavelength only. The data of Sawicki \etal
1997 is shown as filled triangles for 0.28 (0.3) and open triangles for
0.44 (0.45) $\mu$m. The points in the highest redshift bin are shown as
lower limits due to obscuration as discussed above. The data seem to be in
relatively good agreement between the different observations. 
 The figures also show the results of the calculated
luminosity density for each of the respective models, 1a, 1b, 1c, 2,
and the infall models (Timmes \etal 1995, Tosi 1996).
 The SFRs we have chosen, are well defined up to a normalization. 
Rather than normalize the individual SFRs, we 
 can normalize the resultant
luminosity density. Thus, in each of the models, we have normalized ${\cal
L}(\lambda,z)$ to fit the observations of Lilly
\etal (1996) at $z = 0.35$ at $\lambda = 0.44 \mu$m.  After making this
single normalization, the slopes of 
${\cal L}(\lambda,z)$ with respect to $z$ for each wavelength band
as well as the relative magnitudes of ${\cal L}(\lambda,z)$ with
respect to the different wavelengths is a prediction of the model.
As is readily seen from the figures, our calculation of 
${\cal L}(\lambda,z)$ is highly sensitive to the chosen SFR.

The figures show the evolution of the luminosity density from a 
redshift of 5 to the present (if galaxies
 or their progenitors form at $z_{\rm max} \le  5$, a cut should be
made at this redshift).
  For the Einstein-de Sitter Universe
($q_o=0.5$), this corresponds to times from $t \simeq 1$ to 14 Gyr (for
$q_o = 0.1$, it is 1.4 -- 14 Gyr). Whereas the bulk of the data which
exists for
$z < 2$, corresponds to times $t \simeq 2.7$ -- 14 Gyr (3.6 -- 14 Gyr for
the
$q_o = 0.1$ model).

  In model 1b, the SFR is proportional 
to gas mass (Scully \etal 1997). 
The same is true for the late time behavior of model 1a,  when the
``normal" mode is dominant. In these models the gas mass fraction
changes by about a factor of 5 to 10 over the age of the galaxy.
This factor is particularly sensitive  to 
the present gas mass fraction chosen and is not precisely known.
Because of the proportionality between the SFR and the the gas mass
fraction, the change in the SFR in these models is also uncertain.
However, it is not simply the net change in the SFR that is important
when trying to match this high redshift data. Even in
case 1a, where  the SFR decreases by a factor of about 10, we
do not fit the multicolor data and this shows the extreme
sensitivity of the results to the exact form of SFR. In both of these
cases (1a and 1b), it is clear that the slope of ${\cal L}(\lambda,z)$ vs
$z$ is too small. While the increase in the luminosity density is
sufficient (at least in model 1a), the increase occurs over the redshift
range 0 -- 5, rather than 0 -- 2 as indicated by the data. Even more
troublesome is the relative luminosity at the different wavelengths. 
The models are too blue, particularly the bimodal model 1a.  None of
these models can be considered good candidates for cosmic chemical
evolution.  Results for these cases are shown in Figures 2 and 3.

Most problematic as candidates for cosmic evolution models are the
constant SFR (model 1c) and the infall models compiled in Tosi (1996).
As one can see from Figure 1, these models show very little variation in
the SFR and as a result the evolution of the luminosity density does not
even come close to matching the data as seen in Figures 4 and 5.
If one accepts these models as galactic evolution models, then one must
conclude that spiral galaxies such as our own have a star formation history
which is not typical of the light  producing objects observed at high
redshift.  We can not exclude this possibility.

In contrast, model 2, the sequential model with a second mode of the form
 $e^{(-t/\tau)}$ with $\tau$ = 2.5 Gyr, gives a good
 fit both in terms of the slope of
the luminosity densities with respect to $z$ and in color as seen in
Figure 6. The SFR adopted by Madau \etal (1996, 1997) can also be
characterized by  an exponential with a time-scale
$\tau \simeq 1.8$ -- 3 depending on which of their models is used and
the assumed age of the Universe. Of course, these models were designed to
closely reproduce the high redshift luminosity density and we do
not duplicate their results here. The infall model of 
 Timmes \etal (1995) uses a SFR proportional to the square of the gas mass
 fraction and also leads to a fairly sharply decreasing SFR, as indicated
by their
 evolution of SN rate (cf. their fig. 39).  While this model does
somewhat better than the linear infall models (T), it would be hard to
argue that it provides a good fit to the data. The result for the TWW
model is shown in Figure 7. 
It is worth noting that the extragalactic background light resulting from our
 model II will be essentially in agreement with observations since it is 
 similar to that calculated by Madau (1997, see his fig 5)

It is also interesting to compare the derived SFR$(z)$ with the space
density evolution of quasars. After an initial steep rise to $z=1.5$ the
space density flattens and then declines gradually beyond $z=2$ (Hawkins
and Veron 1996, Schmidt \etal 1995). Changes in the space density of 
 quasars may provide important clues to the epoch of the galaxy formation
and related questions. The broad similarities between the SFR in galaxies
and the quasar evolution rate suggest a scenario in which nuclear
starbursts (taking place in elliptical galaxies triggers the quasar
phenomenon (Boyle and Telervich 1998).

\section{Overall metallicity and D destruction}

Element abundances measured in absorbing quasar systems span a large
redshift range (1 to about 4). The various clouds sampled also
differ by their column
 densities; DLAs ($10^{20}$ cm$^{-2}$), Lyman forest
 systems ($10^{14}$ -- $10^{15}$ cm$^{-2}$). We can ask whether or not the
objects observed at high redshift are able to 
 place constraints on the early phases of evolution as calculated by
simple galactic (sic cosmic) models of the type discussed above.  DLAs 
are generally considered
 as precursors of present day disk galaxies (e.g. Wolfe \etal 1995). 
The nature of the protogalactic clumps giving rise to DLAs, however, is
still controversial and their  morphology remains uncertain. The crucial
question is whether  DLAs represent a population of already assembled
proto-disks, or whether
 they are still subgalactic fragments in the process of hierarchical assembly.
 The large HI column densities of DLAs (Lanzetta \etal 1995, Storrie
Lombardie
 \etal 1996) are reminiscent of present day 
galactic disks, but these column densities together with their complex
 line profiles can be
 equally well reproduced by gas-rich merging 
protogalatic clumps with masses expected
 from CDM  hierarchical structure formation models.
 Thus, the large rotating 
 disk hypothesis, which has been favored up to now, can be questioned
(Haehnelt \etal 1997).

 Recent abundance measurements in DLAs  shed new light on
 the problem (Lauroesch \etal 1996 and references therein).  
 The metal abundances (Fe/H, Zn/H) in DLAs at high $z$ are somewhat lower
 than expected for the galactic halos at similar ages but moreover 
 much more dispersed ( Lu \etal 1996, 
 1997, Pettini \etal 1997a,c, Vladilo 1997).
  Moreover,
 the large abundance spread (by up to 2 orders of
 magnitude) at the same redshift seems to indicate that we are dealing with
 objects of different morphologies at different stages of their chemical
evolution or that we are dealing statistically with the
 outer parts  or numerous small systems of low metallicity compared to the
 average metallicity of well formed spirals (Phillips \& Edmunds 1996).
 It is possible that this great dispersion reflects a stochastic phase
 of star formation which is limited to the very early stages of the cosmic
 evolution due
 to the low number and  short lifetime of the stars involved or 
external fringes of disk galaxies not typical of the whole.
 As such, these systems may not represent the true averaged metallicity
to be compared with a cosmic evolutionary model.

 At $z$ greater than 4 , the metallicity of DLAs
 levels off 
 at $[Fe/H]$   = -2 to -2.5. This ``plateau" is identical within
uncertainties
 (which can be as much as a factor of ten) to the metallicity inferred
from CIV absorption
 lines associated with the Lyman forest clouds (Cowie \etal 1995, Tytler
\etal
  1995, Songaila and Cowie 1996). It is worth noting that at $z$ greater
than 3
 there is a
 rapid decline in the space density of quasars (Schmidt \etal 1995). Then,
 at those high redshifts, the formation of structured objects able to trigger
 intense star formation with associated nucleosynthesis has not started or
 has just begun. Thus our models should be limited to $z$ less than
3 -- 4.   
 To conclude, $z$ = 4 seems to be a transition epoch for cosmic chemical
 evolution.

 It has been suggested that intracluster gas is a more appropriate
  sample with which
 evolutionary calculations can be compared (Renzini 1997, Mushotzky and 
 Loevenstein 1997). In Renzini (1997) following Madau \etal (1996), it was
shown that adopting $H_o= 50$, a
 baryonic  density $\Omega_B  = 0.05$ and the density of luminous matter
$\Omega_{lum} = 0.0036$, one obtains a fraction of luminous baryons of
0.07 which is comparable to that
 obtained (6 -- 10 \%) in clusters of galaxies whereas the overall 
metallicity derived from these figures is 7 \% solar compared to the 
 metallicity of clusters of galaxies which is about a third  solar. 
Renzini (1997)
 proposes two possible explanations of cluster- field differences :
 ram pressure stripping (which he later discards) or a
flatter IMF in clusters relative to field galaxies. 
  
 Indeed, $\Omega_B   = 0.05$ corresponds to an intermediate primordial 
deuterium abundance, D/H $\sim 6 \times 10^{-5}$ high compared to the
lower observed abundances measured in quasar absorbers (Tytler, Fan \&
Burles 1996, Burles \& Tytler 1996). Of course, higher values have
 also been published, D/H $\simeq 1 - 2 \times 10^{-4}$ (see e.g.
Vidal-Madjar, Ferlet \&
 Lemoine 1997 for a recent review) yielding $\Omega_B \simeq 0.02$ for
$H_o = 60$.
 In this case, the global metallicity of the Universe would 
be about 0.2 solar which is very close to the observed metallicity of the
clusters of galaxies (about a third solar) and thus
  it would not be necessary to argue that galaxies in and
outside of
 clusters behave differently,  or require different IMFs.
The efficiency  of baryon
 conversion calculated with $\Omega_B = 0.02$ is within a factor of two
compared
 to the one derived from clusters. Choosing a higher value of D/H, one
could conclude that the global metallicity
 of the Universe  is close to the  one observed in the clusters. The high 
 D/H has the advantage to lead to a unique metallicity evolution in the
field
 and in the clusters, whereas the low D/H requires the supplementary
 assumption (Renzini 1997) that the IMF is flatter in galactic clusters.
(The lower value of D/H would require even a greater distinction between
these galaxies.) 
 The global abundance is about constant (of the order of
 0.2 -- 0.3 solar)  from $z=0$ to $z = 2$ or 3; this constraint, as
 suggested by Renzini (1997), make sense 
 for the galactic evolutionary models available. The field ellipticals 
 dominate in mass the spirals  as in clusters, at low redshift (Persic
 and Salucci 1992), and it is not surprising that we find similar
enrichments
 in the intergalactic and intracluster medium, due to
 galactic winds triggered in ellipticalls by SNII (e.g.Elbaz \etal 1995).

We turn now to the destruction of deuterium as implied by the various
evolutionary models considered. The amount of deuterium destruction
is of course very model dependent and can range anywhere from a
factor of about 2 to 15 in our Galaxy (Scully \etal 1997).  If the
observations of D/H in quasar absorption systems relax to a single
well defined value, we would indeed have a strong constraint on
galactic chemical evolution models.  For now, we can simply try to
model the different astration factors implied by the existing
observations.  Nevertheless, among the different models investigated here,
a clear trend with respect to the SFR and the luminosity density is
evidenced. Models with a star formation rate decreasing exponentially and
with a relatively short characteristic time  (as in our model 2
 and the two models from Madau \etal 1997, M1 and M2) are favored.
On the contrary,  models with a moderate SFR variation,
proportional to the gas mass fraction, such as our models 1a, 1b, or the
models compiled in Tosi (1996), similar to model T here, 
do not fit the photometric data. Even the TWW model
with a SFR proportional to the square of the gas mass fraction is
problematic. The constant SFR model 1c fares much worse. 

\vskip .3in
\begin{table} [ht]
\centerline{Table 1: Deuterium Destruction Factors}
\begin{center}
\begin{tabular}{|ccc|}  \hline \hline                   
Model   & total destruction factor &  from $z = 2$ \\
\hline 1a & 12.5 & 7.5  \\
1b & 4.7 & 4.2  \\
1c & 1.5 & 1.5  \\
2  & 12.5 & 10  \\ 
M1 & 13.1 & 6.2  \\
M2 & 16.& 9.1 \\
TWW & 2. & 1.5 \\
T  & 2 -- 3 & 2 -- 3 \\
\hline
\end{tabular}
\end{center}
\end{table}
 
Because deuterium is totally destroyed in the star formation process, the
deuterium destruction factor will be very sensitive to the models we have
considered. In the table below, we show the total deuterium destruction
factor, D$_p$/D$_o$, as well as the factor from $z = 2$ to the present, 
D$_{z = 2}$/D$_o$. When we compare these deuterium destruction factors
with our previous results on the high redshift luminosity data, we see
that although the models which fit the photometric data reasonably well
(2, M1, M2) all destroy significant amounts of deuterium, the converse is
not necessarily true.  That is, models which destroy significants amount
of D/H will not automatically fit the high redshift data. The case in
point is models 1a and 2, which each destroy D/H by more than a factor of
10, yet only model 2 fit the high redshift data well.   However, if one
compares the evolution of D/H in these two models (see e.g. Scully 
\etal 1997), one find that the D/H is destroyed later in model 2 than in
model 1a (this can be seen from the table as well).  Thus the high redshift
 luminosity data not only prefers a large deuterium destruction
factor, but the bulk of the destruction should take place at $z \lsim 2$.
We emphasize that these results can not be used to extrapolate a primordial
D/H abundance from observations of D/H in our own galaxy. As we have
indicated earlier, the star formation history in our own galaxy may have
been very different from that of the typical object which account for the
bulk of the observed luminosity density.  However, in those systems, 
we expect that significant amounts of deuterium destruction has occurred
whatever the primordial D/H ratio may be.

\section{Discussion and conclusions}

Recent observations of the luminosity density at high redshift
(Lilly \etal 1996, Madau \etal 1996, Connolly \etal 1997,
 Sawicki , Lin, \& Yee 1997)  are making it possible for the first time
to test models of cosmic chemical evolution. Madau \etal (1997) tested
several models of cosmic chemical evolution by varying the IMF.  Although
they found that flatter IMF (those containing more massive stars) were
preferred, the luminosity densities were not overly sensitive to the IMF. 
In this work, he have primarily considered the sensitivity of the high
redshift luminosity density to the SFR.  We have run the population
synthesis code of Bruzual
\& Charlot (1997) to calculate ${\cal L}(\lambda,z)$ for a wide variety
of SFRs ranging from a constant SFR to ones which are steeply decreasing
exponentials in time.

Indeed, the high redshift observations, are very discriminatory with
respect to the chosen SFR.  Models in which the star formation rate is
proportional to the gas mass fraction (these are common place in Galactic
chemical evolution) have difficulties to fit the multi-color data from
$z = 0$ to 1.  This includes many of the successful Galactic infall models.
In contrast, 
models with  a SFR proportional  to $e^{-t/\tau}$  with $\tau$ between 2
to 4 or to some extent, proportional to $\sigma^2$ are favored. 
Further consequences of the the adopted histories of star formation could
be worked out including a calculation of the
	brightness of the night sky including the FIR (Guiderdoni \etal 
1998).
  
While we can not conclude that all models with large deuterium destruction
factors are favored, it does seem that models which do fit the high redshift data destroy significant amounts of D/H.  On the other hand, we can
not exclude models which destroy only a small amount of D/H as Galactic
models of chemical evolution. In this case, however the evolution of our
Galaxy is anomalous with respect to the cosmic average. If the low D/H
measurements of Tytler, Fan \& Burles (1996) and Burles \& Tytler (1996) 
hold up, then it would seem that our Galaxy also has an anomalously high 
D/H abundance.  That is we would predict in this case that the present
cosmic abundance of D/H is significantly lower than the observed ISM
value.  If the high D/H observations 
(Songaila \etal 1994, 1997, Carswell \etal 1994, Wampler 
1996, Webb \etal 1997) hold up, we would conclude that our Galaxy is
indeed representative of the cosmic star formation history.

We note that our detailed results are somewhat dependent on the chosen
cosmological model.  For example, had we chosen to allow for values of
$\Omega_o < 1$, or a chosen a 
model with with a cosmological constant, $\Lambda$ consistent with the
revised age of  the universe and SNIa observations at moderate redshifts
(Perlmutter \etal 1997)  we could reach somewhat different conclusions.
The effect of lowering $\Omega_o$ and $q_o$ was discussed in Lilly \etal
(1996).  The leading effect is a lowering of the luminosity density data
at high redshift.  For example for $\Omega_o = 0.1$ and $\Omega_\Lambda
= 0$, $\Delta \log {\cal L} = -.43 \log (1 + z)$,
while for $\Omega_o = 0.1$ and $\Omega_\Lambda
= .9$, $\Delta \log {\cal L} = -1.12 \log (1 + z)$.
Such a shift can have dramatic consequences on the comparison of the
model prediction and data.  This effect led Totani, Yoshii, \& Sato (1997)
to conclude that the high redshift data indicated a non-zero
cosmological model.  A similar conclusion would be reached if instead of
varying the SFRs and hence the chemical evolution models, we varied
the cosmological models.  In Figure 8, we show the luminosity density
for model 1b, in the context of a  $\Omega_o = 0.1$ and $\Omega_\Lambda
= .9$ cosmological model.	Now the fit is quite reasonable at the expense
of introducing a cosmological constant.  Thus, several of the models
considered which show only a modest rise in ${\cal L}$ could be made to
better fit the data in this case. While we could certainly state that 1b
is compatible with the data for non-zero $\Lambda$, we could not claim
evidence for a non-zero $\lambda$ on the basis of this model. This
distinction is important. Furthermore, it is difficult to imagine a
greater increase in
$\Lambda$ relative to the one considered, and so it is unlikely that
models 1c or T could be brought into agreement with the high redshift
data.

We have demonstrated that the observations of the luminosity density at
high redshift is a key discriminator among models of cosmic chemical
evolution with different star formation rates.  Future observations of
this type coupled with measurements of D/H in quasar absorption systems
will help us understand not only the average cosmic evolution but also
whether or not our own Galaxy is typical of that average. 

\noindent
{\bf Acknowledgments}

\noindent
We would like to thank Stephane Charlot for running his population
sysnthesis code and for pertinent comments on the text. For all his
support, we acknowledge him very deeply.  We are  also indebted to Sean
Scully for calculations of Deuterium destruction
 in different galactic evolutionary models. 
We are also grateful to the referee for pertinent remarks.
This work was carried out under
 the auspices of PICS 319, CNRS. The work of KAO 
was supported in part by  DOE grant DE--FG02--94ER--40823.

\beginapjbib

\bibitem Allard, F. \& Hauschildt, P.H., 1995, ApJ, 445, 433

\bibitem Alongi,M., \etal, 1993, A \& AS, 97, 851

\bibitem Bessel, M.S., Brett, J., Scholtz, M., \& Wood, P., 1989, A \& AS,
77,1

\bibitem Bessel, M.S., Brett, J., Scholtz, M., \& Wood, P., 1991, A \& AS,
89, 335

\bibitem Boyle, B.J.\& Terlevich, R.J., 998, MNRAS, 293, L49

\bibitem Bressan, A., Fagotto, F., Bertelli, G., \& Chiosi, C. 1993, A \&
AS, 100, 647

\bibitem Bruzual, A.G. \& Charlot, S., 1997, in preparation

\bibitem Bruzual, A.G., \etal, 1997, AJ, 114, 1531

\bibitem Burles, S. \& Tytler, D. 1996, ApJ, 460, 584

\bibitem Carigi, L. 1996, Rev. Mex. A. C., 4, 123

\bibitem Carswell, R.F., Rauch, M., Weymann, R.J., Cooke, A.J. \&
Webb, J.K., 1994, MNRAS, 268, L1 

\bibitem Charlot, S., Worthey, G., \& Bressan, A., 1996, ApJ, 457, 625 

\bibitem  Connolly, A.J. Szalay, A.S., Dickenson, M., SubbaRao, M.U., \&
Brunner, R.J. 1997, ApJ, 486, L11

\bibitem Cowie, L.L., Songaila, A., Kim, T.S. \& Hu, E.M., 1995, AJ, 109, 
1522


\bibitem, Edmunds, M.G. \& Phillips, S. 1997, MNRAS, 292, 733

\bibitem Elbaz, D., Arnaud, M. \& Vangioni-Flam E. 1995, A \& A, 303, 345


\bibitem Fagotto, F., Bressan, A., Bertelli, G., \& Chiosi, C., 1994a, A
\& AS,  105, 39

\bibitem Fagotto, F., Bressan, A., Bertelli, G., \& Chiosi, C., 1994b, A
\& AS,  104, 365

\bibitem Fagotto, F., Bressan, A., Bertelli, G., \& Chiosi, C., 1994c, A
\& AS,  105, 29

\bibitem Fall, S.M. \& Pei, Y.C. 1993, ApJ 402, 479

\bibitem Fluks, M. \etal, 1994, A \& AS, 105, 311

\bibitem Galli, D., Palla, F., Ferrini, F., \& Penco, U. 1995, ApJ, 443,
536


\bibitem Geiss, J. 1993, in {\it Origin
 and Evolution of the Elements} eds. N. Prantzos,
E. Vangioni-Flam, and M. Cass\'{e}
(Cambridge:Cambridge University Press), p. 89

\bibitem Girardi, L. \etal, 1996, A \& AS, 117, 113

\bibitem Guiderdoni, B. \etal 1998, MNRAS, 295, 877

\bibitem Hawkins, M.R.S. \& Veron, P. 1996, MNRAS, 281, 348

\bibitem Haehnelt, G., Steinmetz, M., \& Rauch, R., preprint
astro-ph/9706201


\bibitem Iglesias, C.A.J., Rogers, F.J., \& Wilson, B.G., 1992, ApJ, 397,
717


\bibitem Kurucz, R.L.,1992, in IAU Symp, 149, The Stellar Population of
 Galaxies, ed. B. Barbuy \& A. Renzini (Dordrecht: Kluwer), 225

\bibitem Lanzetta, K.M. Wolfe, A.M. \& Turnshek, D.A. 1995, ApJ, 440, 435

\bibitem Larson, R.B. 1974, MNRAS, 169, 229

\bibitem Larson, R.B. 1986, MNRAS, 218, 409

\bibitem Lauroesch, J.T., Truran, J.W., Welty, D.E. \& York, D., G. 1996,
 PASP, 108, 641

\bibitem Lejeune,T., Cuisinier, F., \& Buser, R., 1997,  A \& AS, 125, 229

\bibitem Lilly, S.J., Le Fevre, O., Hammer, F., \& Crampton, D.  1996,
ApJ, 460, L1

\bibitem Linsky, J.L., Brown, A., Gayley, K., Diplas, A., Savage, B. D.,
Ayres, T. R., Landsman, W., Shore, S. N., \& Heap, S. R. 1993, ApJ, 402,
694

\bibitem Linsky, J.L.,  Diplas, A., Wood, B.E.,  Brown, A.,
Ayres, T. R., \& Savage, B. D., 1995, ApJ, 451, 335

\bibitem Loveday, J. Peterson, B.A., Efstathiou, G. \& Maddox, S.J. 1992,
ApJ, 390, 338

\bibitem Lu, L., Sargent, W.L.W., Barlow, T.A., Churchill, C.W., \&
 Vogt, S.S. 1996, ApJS, 107, 475

\bibitem Lu, L. Sargent, W.L.W., \& Barlow, T.A. 1997, preprint
astro-ph/9710370

\bibitem Madau, P., 1997, preprint astro-phys 9709147

\bibitem Madau, P., Ferguson, H.C.,  Dickenson, M.E.,  Giavalisco,
M., Steidel, C.C., \& Fruchter, A. 1996, MNRAS, 283, 1388

\bibitem Madau, P., Pozzetti, L., \& Dickinson, M., 1997, astro-ph/9708220

\bibitem Matteuchi, F. \& Francois, P. 1989, MNRAS, 262, 545

\bibitem Meurer, G.R. 1997, preprint, astro-ph/9708163

\bibitem Mushotzky, R.F. \& Loevenstein, M., 1997, ApJ, 481, L63

\bibitem Pei, Y.C. \& Fall, S.M., 1995, ApJ, 454, 69

\bibitem Persic, M. \& Salucci, P. 1992, MNRAS, 258, 14p

\bibitem Pettini, M. \etal , 1997a, ApJ, 478, 536

\bibitem Pettini, M. \etal, 1997b, preprint astro-ph/9707200

\bibitem Pettini, M. \etal, 1997c, ApJ, 486, 665

\bibitem Perlmutter, S. \etal, 1997, ApJ, 483, 565

\bibitem Phillips, S. \& Edmunds,M.G. 1996, MNRAS, 281, 362

\bibitem Prantzos, N. 1996, A\&A, 310, 106

\bibitem Renzini, A., 1997, ApJ, 488, 35

\bibitem Savaglio, S. 1997, preprint, astro-ph/9709154

\bibitem Sawicki, M.J., Lin, H., \& Yee, H.K.C. 1997, AJ, 113, 1 

\bibitem Scalo, J., 1986, Fund. Cosm. Phys. 11, 1

\bibitem Schmidt, M., Schneider, D.P. \& Gunn, J.E., 1995, AJ, 110, 68 

\bibitem Scully, S.T., Cass\'{e}, M., Olive, K.A., Schramm, D.N., 
Truran, J., \& Vangioni-Flam, E. 1996, ApJ, 462, 960

\bibitem Scully, S.T., Cass\'{e}, M., Olive, K.A., \& Vangioni-Flam, E.
1997, ApJ, 476, 521

\bibitem Soifer, B.T. \& Neugebauer, G., 1991, AJ, 101, 354

\bibitem Songaila, A., Cowie, L.L., Hogan, C. \& Rugers, M. 1994,
Nature, 368, 599

\bibitem Songaila, A., Wampler, E.J. \& Cowie, L.L., 1997, Nature, 385, 137

\bibitem Songaila, A. \& Cowie, L.L., 1996, AJ, 112, 335

\bibitem Steidel, C.C., Giavalisco, M., Pettini, M., Dickenson, M. \&
Adelberger, K. 1996, ApJ, 462, L17

\bibitem Storrie-Lombardi, V., Mc Mahaon, R.G. \& Irwin, M.J., 1996,
MNRAS, 283, L79
 
\bibitem Timmes, F.X., Woosley, S.E., \& Weaver, T.A. 1995, ApJSupp,
98, 617

\bibitem Tinsley B.M. 1980, Fund. Cosmic Phys., 5, 287

\bibitem Tosi, M., 1996, in "From Stars to Galaxies", ASP, Conference
Series, vol.98, 299

\bibitem Totani, T., Yoshii, Y. \& Sato, K., 1997, ApJ, 483, L75

\bibitem Tytler, D. \etal , 1995, in "QSO Absorption Lines", Edts G. Meylan
 (Springer-Verlag), p. 289 

\bibitem Treyer, M.A., Ellis, R.S., Milliard, B. \& Donas, J., 
1997, in The Ultraviolet Universe at Low
 and High Redshift, Edts W. Waller, (Woodbury:AIP Press), in press
 astro-ph/9706223

\bibitem Tytler, D., Fan, X.-M., \& Burles, S. 1996, Nature, 381, 207

\bibitem Vader, P. 1986, ApJ, 305, 669


\bibitem Vidal-Madjar A., Ferlet R., \& Lemoine, M. 1997, in "Structure and
Evolution
 of the Intergalactic Medium from QSO Absorption Line Systems", eds. P.
 PetitJean, S. Charlot, Frontiere, p. 355

\bibitem Vladilo, G., 1997, preprint astro-ph/9710026

\bibitem Wampler E.J., Nature, 383, 308

\bibitem Webb, J.K. \etal. 1997, Nature, 388, 250 

\bibitem Wolfe \etal 1995, ApJ, 454, 698

\endapjbib

\newpage

\section*{Figure Captions}

{\bf Figure 1:} 
A comparison of the various SFR histories considered.
Model 1a (from Scully \etal 1997, shown as a dashed curve) has bimodal
star formation, the late time behavior of the SFR is proportional to the
gas mass; Model 1b (from Scully \etal 1997, shown as a solid curve) is
more standard with the SFR also proportional to the mass in gas; Model 1c
(from Scully \etal 1997, shown as a dotted curve) has a constant SFR;
Model 2 (from Scully \etal 1997, shown as a dot-dashed curve) is a
sequential model with a late time behavior given as 
$\psi(t) \propto exp(-t/2.5)$; Model M1 (from Madau \etal 1997, shown as
a thick solid curve) and Model M2 (also from Madau \etal 1997, shown as a
thick dot-dashed curve) were both chosen to fit the high redshift
photometric data and both have a late time behavior which is well
characterized by an decreasing exponential; Model T  (shown as a thin
solid curve) is a SFR representative of the infall models compiled in Tosi
(1996); Model TWW (from Timmes
\etal 1995, shown as a thin solid curve) has a 
SFR proportional to the square of the gas mass fraction.

\noindent{\bf Figure 2:} .
The tricolor luminosities densities (UV, B and IR) at $\lambda = 0.28,
0.44 and 1.0 \mu$m, in units of (h/.5) WHz$^{-1}$Mpc$^{-3}$ as a function
of redshift for model 1a. The data are taken from Lilly \etal (1996)
(filled squares), Connolly \etal (1997) (open circles), and Sawicki \etal
(1997) (open sqaures).

\noindent{\bf Figure 3:}  Same as Figure 2 for model 1b.

\noindent{\bf Figure 4:} Same as Figure 2 for model 1c.

\noindent{\bf Figure 5:} Same as Figure 2 for model T.

\noindent{\bf Figure 6:} Same as Figure 2 for model 2.

\noindent{\bf Figure 7:} Same as Figure 2 for model TWW.

\noindent{\bf Figure 8:} Same as Figure 3 for a cosmological model with
$\Omega_o = 0.1$ and $\Omega_\Lambda = 0.9$.


 \newpage

\begin{figure}[htb]
\vskip 1in
\hskip 1in
\epsfysize=7truein
\epsfbox{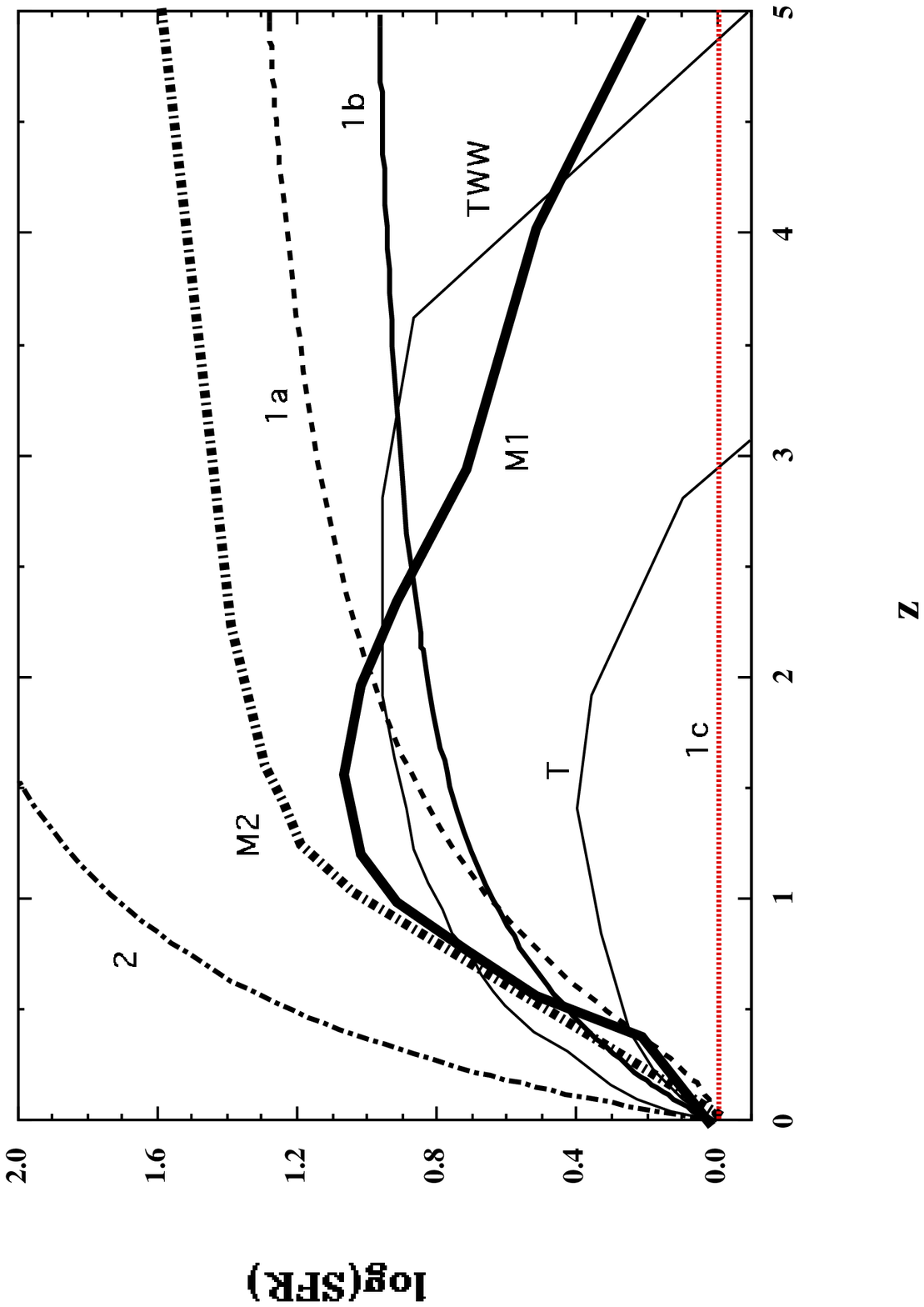}
\end{figure}  
\newpage

\begin{figure}[htb]
\epsfysize=7.5truein
\epsfbox{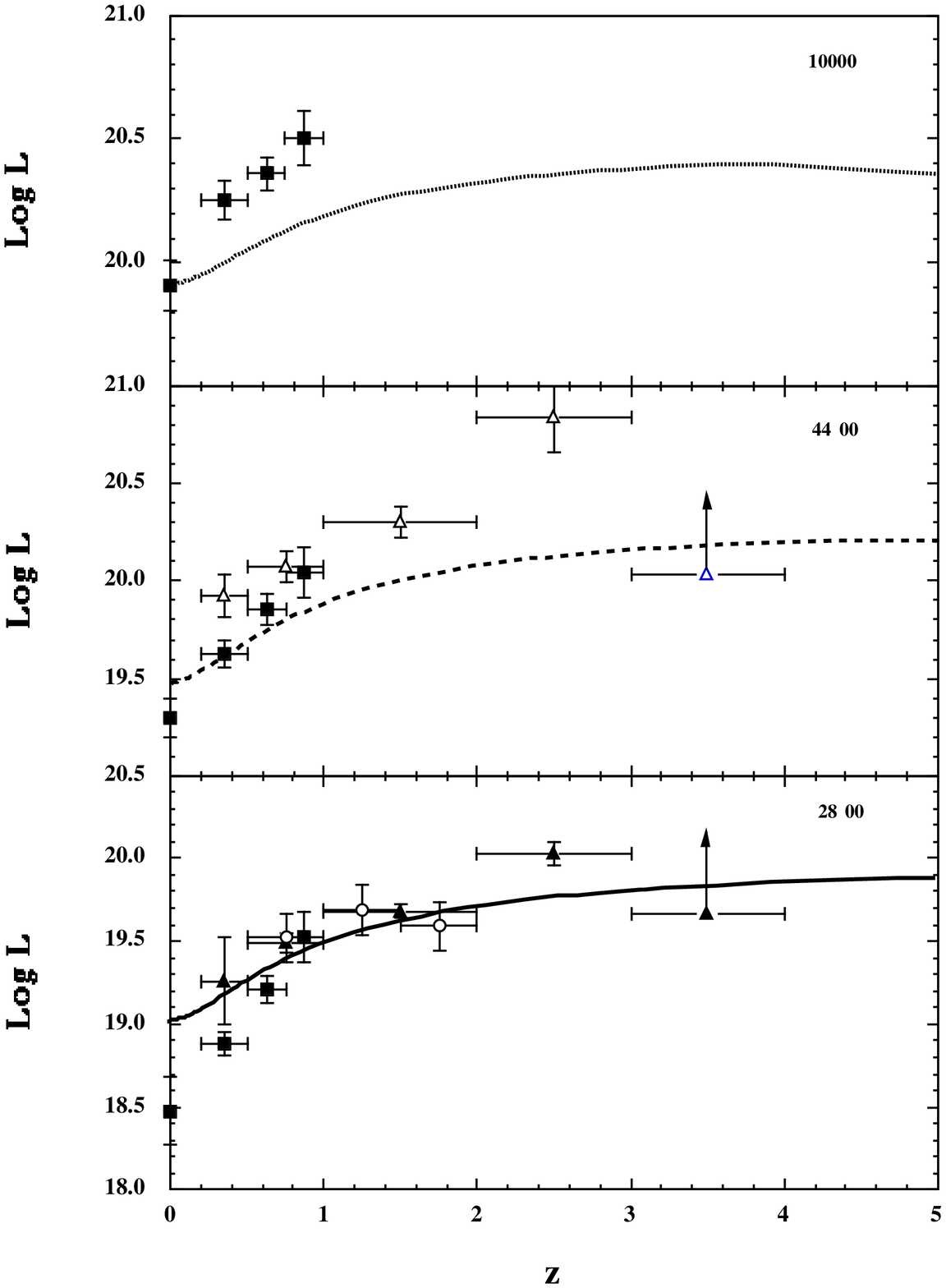}
\end{figure}  

\newpage

\begin{figure}[htb]
\epsfysize=7.5truein
\epsfbox{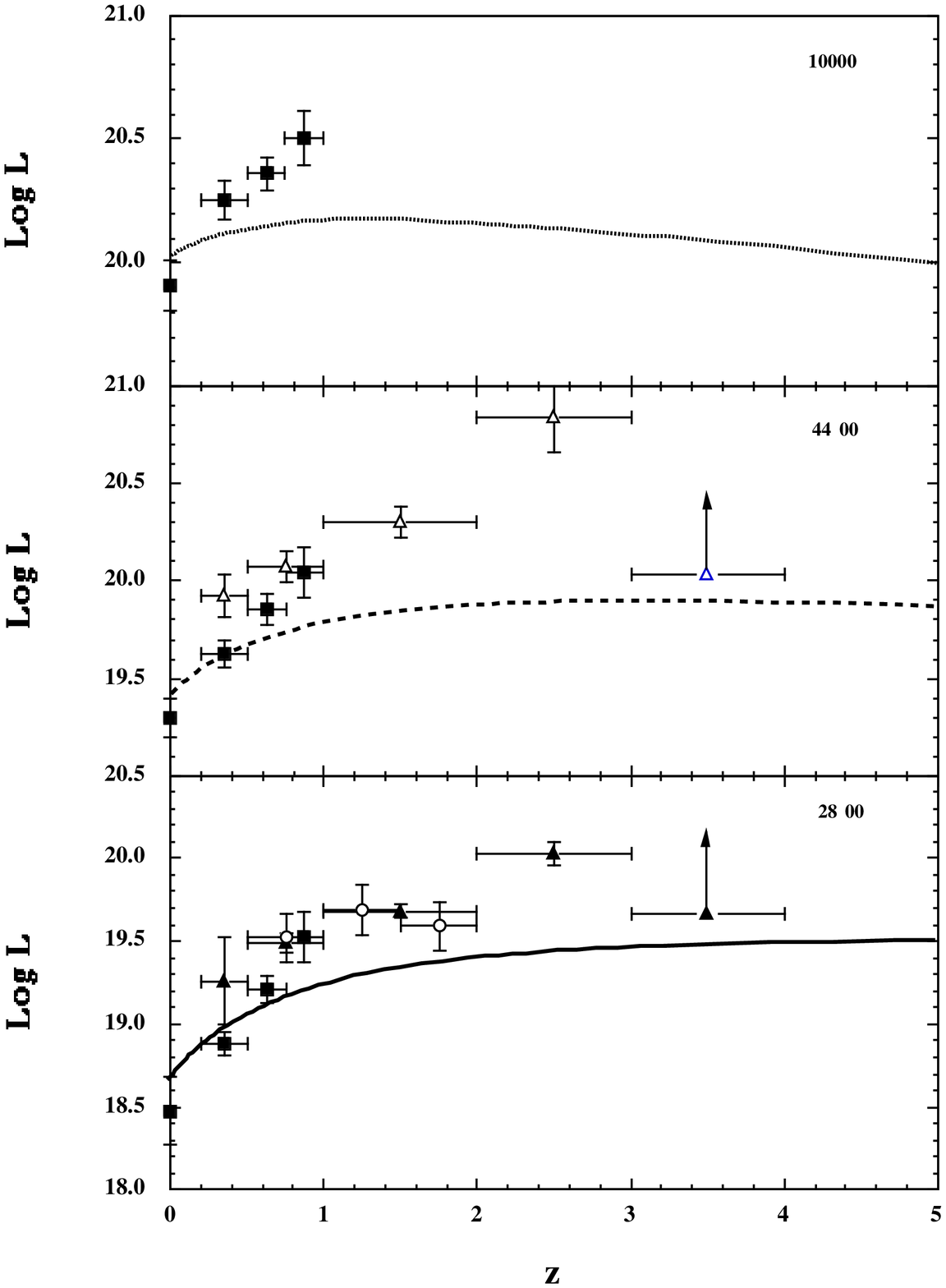}
\end{figure}

\newpage

\begin{figure}[htb]
\epsfysize=7.5truein
\epsfbox{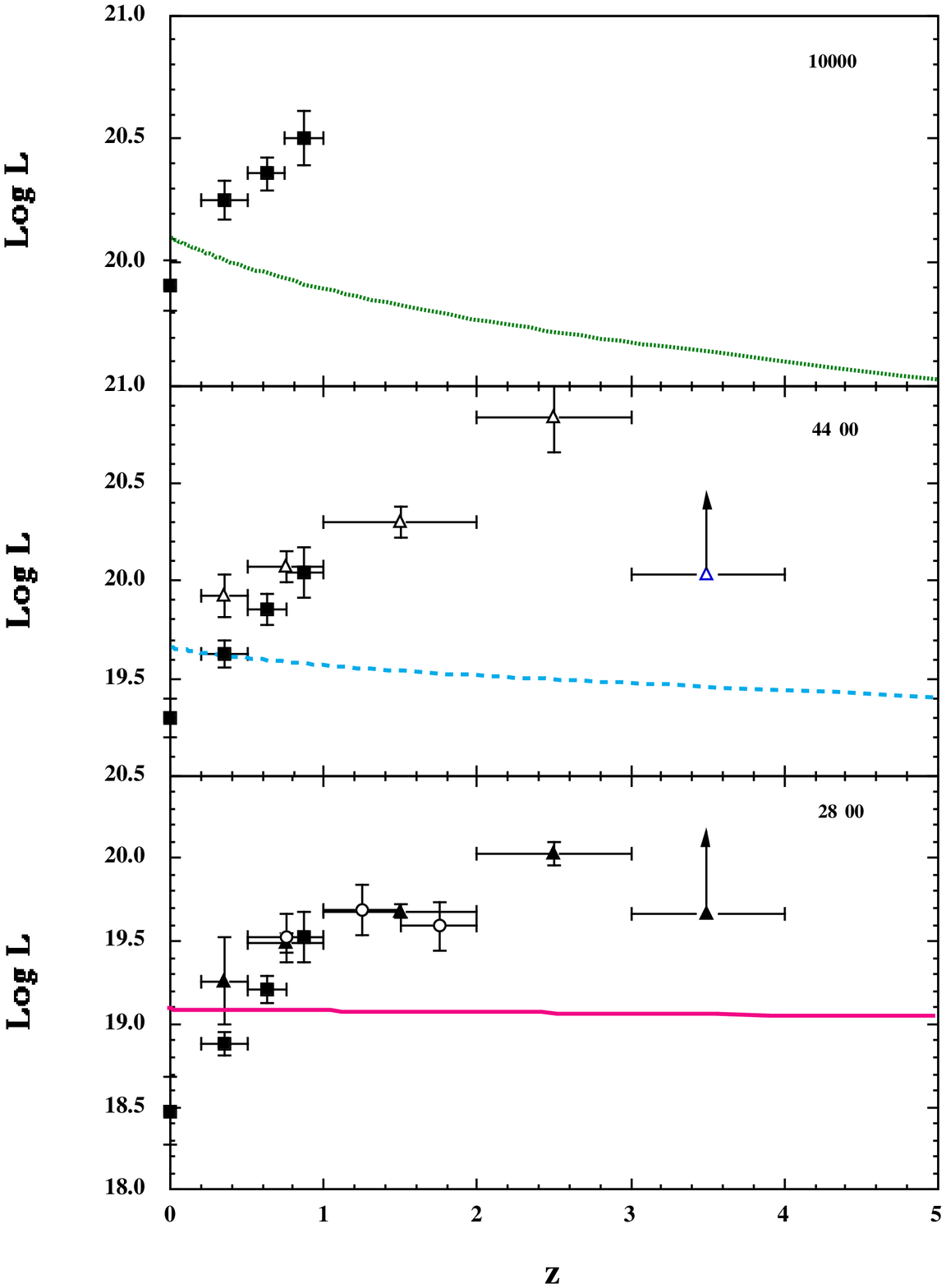}
\end{figure}

\newpage

\begin{figure}[htb]
\epsfysize=7.5truein
\epsfbox{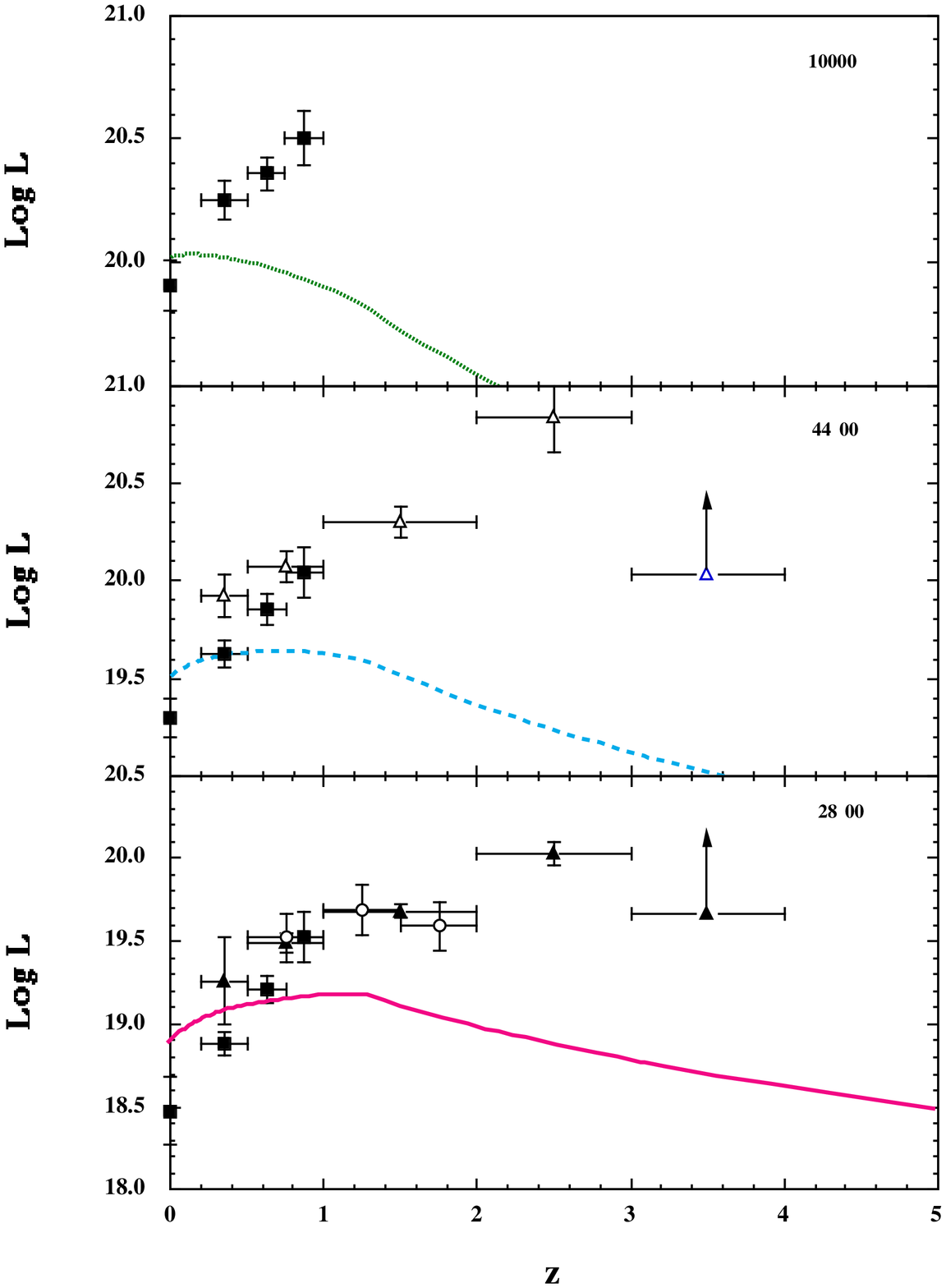}
\end{figure}

\newpage

\begin{figure}[htb]
\epsfysize=7.5truein
\epsfbox{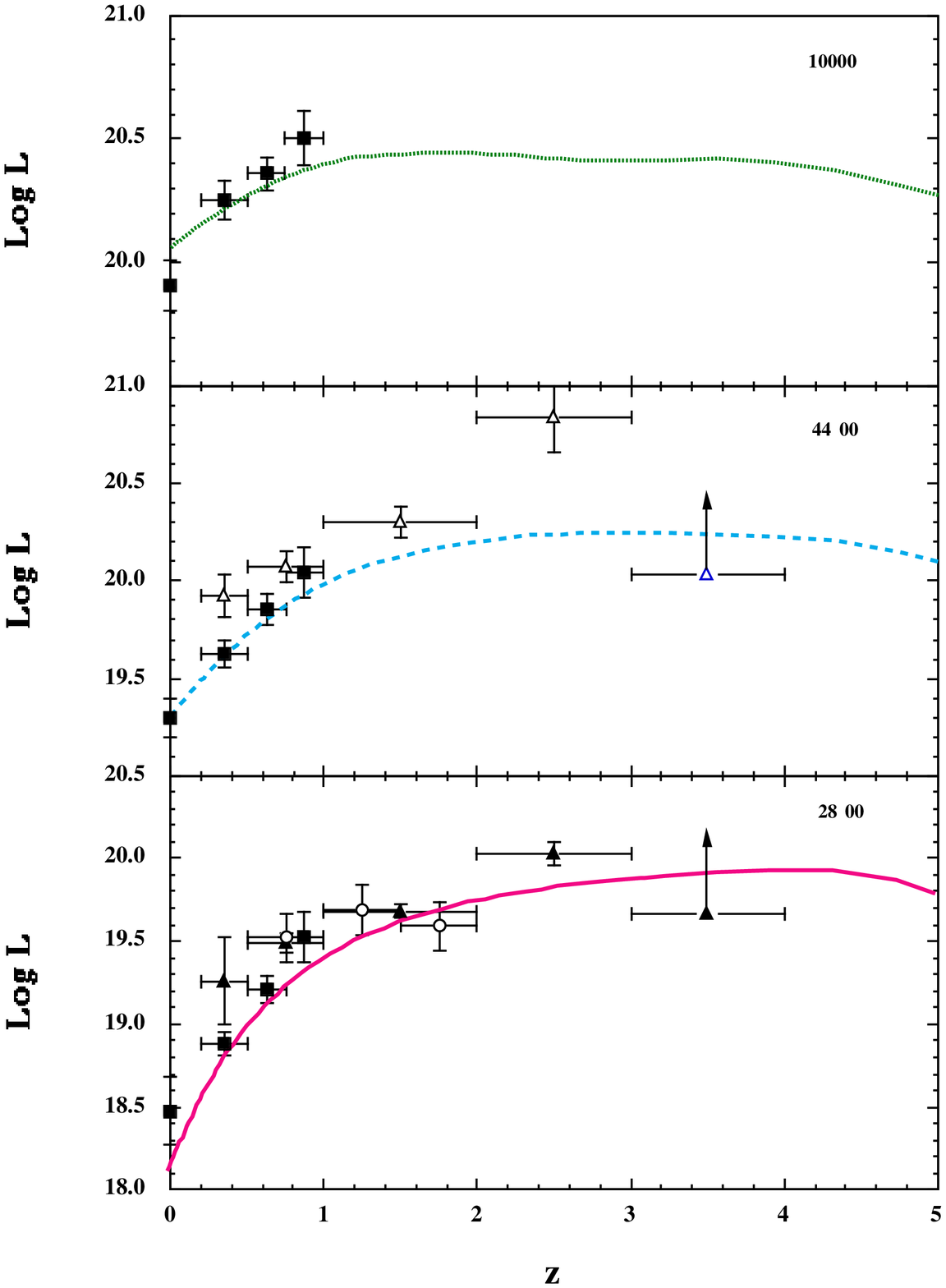}
\end{figure}

\newpage

\begin{figure}[htb]
\epsfysize=7.5truein
\epsfbox{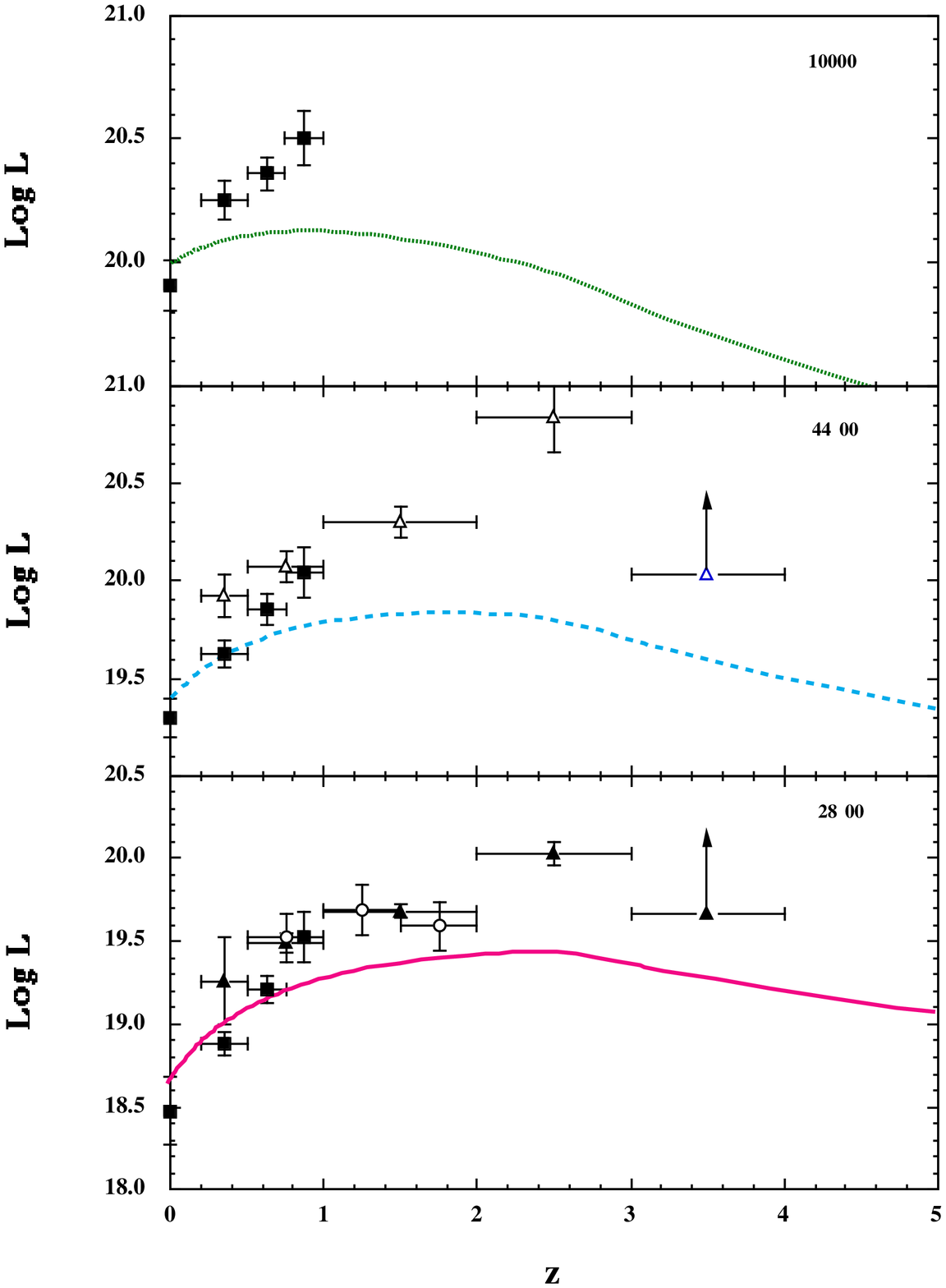}
\end{figure}

\newpage

\begin{figure}[htb]
\epsfysize=7.5truein
\epsfbox{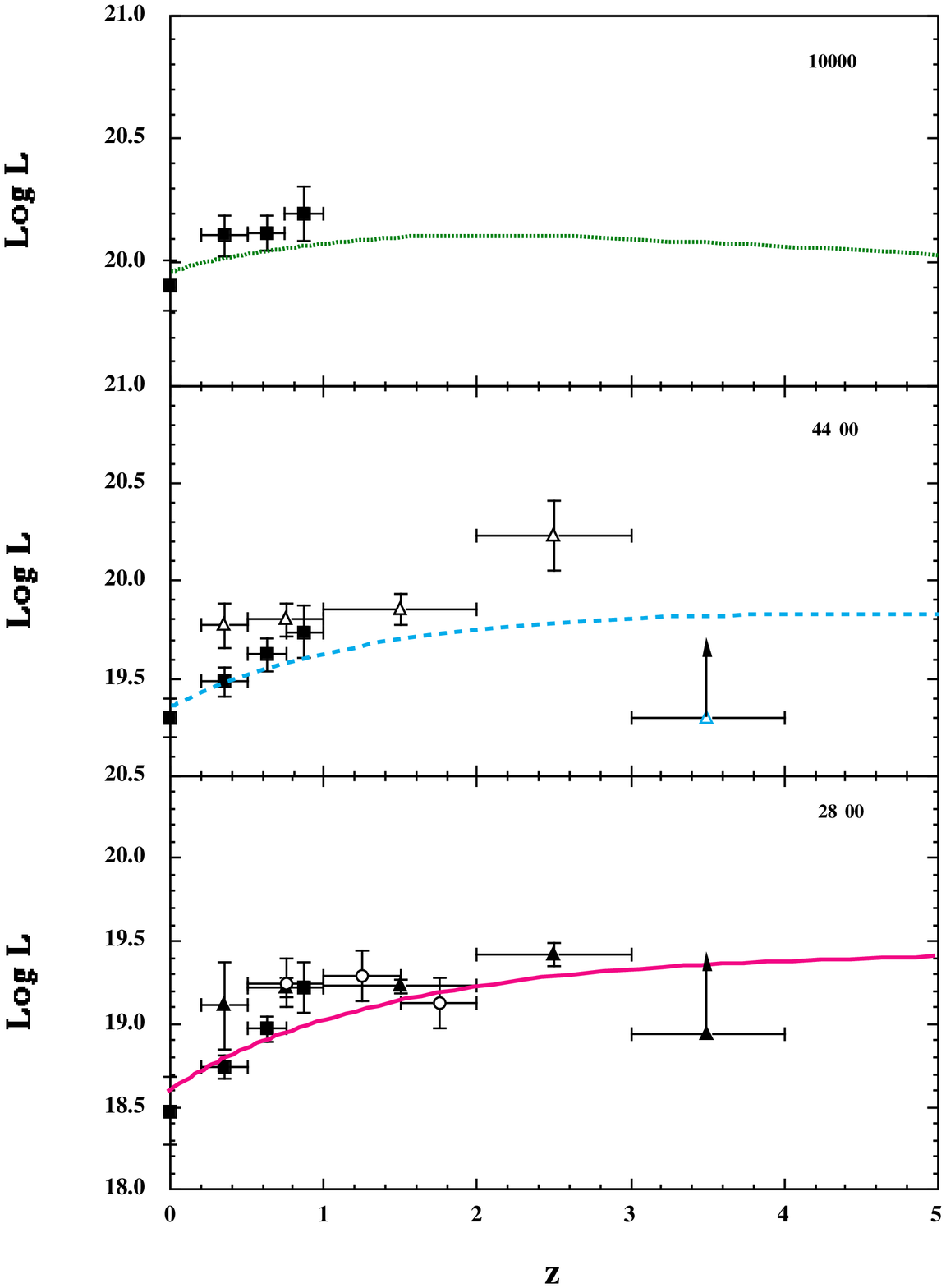}
\end{figure}

\end{document}